\documentclass[twocolumn,showpacs,pre,preprintnumbers,amsmath,amssymb,aps]{revtex4}
\usepackage{dcolumn}
\usepackage{graphicx}
\usepackage{bm}
\begin{document}

\title{Correlation function and generalized master equation of arbitrary age}
\author{Paolo Allegrini$^{1}$, Gerardo Aquino$^{2}$, Paolo Grigolini$^{2,3,4}$,
Luigi Palatella$^{5}$, Angelo Rosa$^{6}$ and Bruce J. West$^{7}$}
\affiliation{$^{1}$INFM, unit\`a di Como, Via Valleggio 11, 22100 Como, Italy}
\affiliation{$^{2}$Center for Nonlinear Science, University of North Texas,
P.O. Box 311427, Denton, Texas 76203-1427}
\affiliation{$^{3}$Dipartimento di Fisica dell'Universit\`{a} di Pisa and
INFM,
Via Buonarroti 2, 56127 Pisa, Italy }
\affiliation{$^{4}$Istituto dei Processi Chimico Fisici del CNR Area 
della Ricerca di Pisa, Via G. Moruzzi 1, 56124 Pisa, Italy}
\affiliation{$^{5}$Istituto dei Sistemi Complessi del CNR, 
P.le A. Moro 2, 00185 Rome, Italy}
\affiliation{$^{6}$Institut de Math{\'e}matiques B, Facult{\'e} 
des Sciences de Base, {\'E}cole Polytechique F{\'e}d{\'e}rale de 
Lausanne, 1015 Lausanne, Switzerland}
\affiliation{$^{7}$Mathematics Division, Army Research Office, Research 
Triangle Park, NC 27709, USA}

\begin{abstract}
We study a two-state statistical process with a non-Poisson distribution of sojourn times. 
In accordance with earlier work, we find that this process is characterized by aging and we study three different ways to define the correlation function of  arbitrary age of the corresponding dichotomous fluctuation.  These three methods yields exact expressions, thus coinciding with the recent result by
 Godr\`{e}che and Luck [J. Stat. Phys. 104, 489 (2001)].  The first method rests on the lines established in an earlier paper [Allegrini et al., Phys. Rev. {\it E} 68, 056123]. These authors built up an infinitely aged Generalized Master Equation (GME), compatible with the ordinary Onsager Principle. %Herein we generate first a GME of arbitrary 
Herein we generate first a GME of arbitrary age, then we derive from it an expression for the corresponding correlation function, 
based on the Onsager principle extended to a condition of generic age.
 To derive the second exact expression of correlation  function of arbitrary age, we use 
a Liouville-like approach, based on a model mimicking the environment responsible for the fluctuations of the dichotomous variable under study, and producing the same non-Poisson distribution of sojourn times as that considered in the earlier treatment. This exact expression refers to time $t$, while the earlier exact expression refers to the variable $u$, which is the Laplace conjugate of $t$. Finally, the third exact expression, again with respect to time, is derived from the direct observation of the dichotomous sequence generated by the two-state system fluctuations. The resulting expression is implicit, and
we adopt a convenient approximation to
obtain a simple analytical formula. 
This approximation rests on the renewal nature of the process, which resets
to zero the memory of the system after the occurrence of any event, thereby
suggesting that a Markovian approximation can be made. Actually, non-Poisson
statistics yields infinite memory at the probability level, thereby breaking
any form of Markovian approximation, including the one adopted herein. 
For this reason, we check the accuracy of the analytical formula by comparing it with the numerical treatment of the second of the three exact expressions. We find that, although not exact, the simple analytical expression for the correlation function of arbitrary age
 is very accurate. 

\end{abstract}


\maketitle

\section{Introduction}
\vspace{0.3cm}
\label{intro} The phenomenon of aging has been known for a long time as a
property of spin glasses and polymers \cite{aging}. Part of the reason for
the more recent interest in this phenomenon has to do with the predicted
breakdown of certain fundamental assumptions made in equilibrium statistical
mechanics when applied to strongly disordered systems. For example, the
Onsager Principle \cite{onsager}, being the relaxation of a perturbed system back to its
equilibrium state as described by an unperturbed autocorrelation function,
is violated in anomalous diffusion and anomalous relaxation. More recent papers on
this phenomenon are devoted to studying aging in diffusion processes
occuring in $d$-dimensional lattices \cite{cecile}, in low dimensional
environments \cite{laloux} and in the quantum dynamics of dissipative free
particles \cite{pottier}. Most recently  \cite{barkai,gerardo} there has been some interest in the
manifestation of aging in processes described by means of the Continuous
Time Random Walk (CTRW) formalism \cite{montroll}.

A natural way of formally expressing aging is through the correlation
function of the stochastic variable of interest, $\xi $, which is expressed
as an ensemble average indicated by the brackets

\begin{equation}
\frac{\langle \xi (t_{1})\xi (t_{2})\rangle _{t_{a}}}{\langle \xi
^{2}\rangle }\equiv \Phi _{\xi }^{(t_{a})}(t_{1}-t_{2}).  \label{aging}
\end{equation}
The averaging brackets carry a subscript stressing that this is a $t_{a}$-old property, rather than the traditional aged property,
 only depending on $|t_{1}-t_{2}|$. The time difference indicates the stationarity of the
underlying process, but the $t_{a}$-superscript denotes a dependence on
the age of the process as well. In the case of a dichotomous variable obeying renewal theory,
the exact expression for the Laplace transform of this age-dependent correlation
function was found by Godr\`{e}che and Luck \cite{luck}. Their exact result
was recently recovered by Margolin and Barkai \cite{margolin} as a special
case of a more general expression, since the latter authors do not require
the condition that the two states of the dichotomous variable $\xi $ have
the same waiting-time distribution. However, herein we make the same
assumption as did Godr\`{e}che and Luck \cite{luck}, and derive their result
along the lines of the recent work of Ref. \cite{gerardo}. Allegrini et al. 
\cite{gerardo} noticed that in the non-Poisson case the well known
Generalized Master Equation (GME) of Kenkre, Montroll and Shlesinger \cite
{GME} becomes incompatible with the Onsager principle \cite{onsager} and
found a way to make the GME compatible with the aged condition. However,
they \cite{gerardo} left unsolved the problem of deriving a GME of arbitrary
age, which is equivalent to making the Onsager principle compatible with an
incompletely aged system. Here we solve this problem and establish that this
solution leads to the exact expression for the correlation function of arbitrary age
found by Godr\`{e}che and Luck \cite{luck}.

The GME derived herein does not have the same origin as those widely
discussed in the literature \cite{zwanzig}. The Zwanzig GME, in fact, one of
the most popular master equations, is derived from a first principles
procedure, starting from the statistical Liouville equation of the whole
universe. The Zwanzig GME is a projection of this universal Liouville
equation onto the Hilbert space of the system of interest. The approach used
to derive the GME herein is quite different from that of Zwanzig and is
based on the experimental observation of non-Poisson dichotomous signals.
Examples of such signals include those produced by an ionic channel \cite
{west} and by blinking quantum dots \cite{blinking}. We build up a GME that
is compatible with the experimentally determined non-Poisson nature of these
processes, assuming the applicability of renewal theory. We leave open the
question as to the source of randomness, but note in passing that the
fluctuations in the system variable are generated by the environment. A simplified dynamical model has reproduced the essential statistical properties of the system of interest, in the case of infinitely aged GME \cite{paper1}. In the case of arbitrary age,  not only a first principle derivation is missing, a condition shared by the infinitely aged case: there is, to date, not even a  heuristic derivation derivation of the corresponding GME. Herein we provide a heuristic derivation of a  GME  with arbitrary age, one based on an exact expression for the corresponding distribution of sojourn time, and we use this result to derive an exact expression for the correlation function.
%A simplified dynamical model has reproduced the essential statistical
%properties of the system of interest \cite{paper1}, however, there is, to
%date, no first principles derivation of the age-dependent GME.  Herein we provide a heuristic derivation of the age-dependent GME, one based on an exact expression for %the age-dependent distribution of sojourn time, and we use this result to derive an exact expression for the correlation  function. 

In conclusion, we point out that we rework the problem from a number of
perspectives. We do this because each approach provides separate and distinct
insights into the phenomenon described by  a GME of arbitrary age. The first
perspective adopted in this paper focuses on the derivation of a GME, based
on the experimental observation of the time evolution of a trajectory,
characterized by rare jumps from one state to another. The second
perspective uses a Liouville-like approach to the time evolution of the
variable of interest and its environment. No attempt is made to derive the
former dynamic picture (trajectory) from the latter (probability density).
The probability density perspective is characterized by infinite memory,
yet, the statistical process under study is generated by random critical
events, whose occurrence erases the memory of earlier events. Therefore, we
think it is prudent to examine, yet again, the same process from a third
perspective, one based on the direct observation of the sequence of rare
random events.

We find that the use of three different perspectives is fruitful in shedding
light on the recent observation made by Sokolov, Blumen and Klafter \cite
{externalfield}. Their results, in our opinion, imply the breakdown of
certain well-established notions of equilibrium statistical mechanics, such
as linear response theory as a prescription for predicting the effect of
externally perturbing a system out of equilibrium.

The outline of the paper is as follows. In Section II we employ the reduced
density perspective, namely, we build up the GME of arbitrary age. In Section III we
study the time evolution of the total distribution density. In Section IV we
use a third method, based on the probability of occurrence of the rare
random events. Section V is devoted to concluding remarks.

\section{The reduced density perspective}

\label{redpersp} This section is devoted to the application of the reduced
density perspective to the construction of the GME. In the literature of
statistical physics this perspective implies a contraction on the total
density onto a prescribed subspace. Consequently, constructing the equation
of motion of the total density matrix should be the first step. The reason
why we reverse the perspective, thereby confining the total density
treatment to the next section is given by the fact that non-Poisson
statistics has the effect of making the ordinary approach extremely
difficult, if not impossible, as discussed in the companion papers 
\cite{paper1,paper2}. The derivation of the GME is made possible by
expressing the higher-order correlation functions in terms of the
second-order correlation function, via simple expressions that are violated
by the Poisson statistics. This is the reason why we derive the GME from the
Continuous Time Random Walk (CTRW) of Montroll and Weiss \cite{montroll, mw}, rather
than from the equation of motion for the total distribution density.

\subsection{From the generalized master equation to the age-dependent correlation
function}

The reduced density matrix perspective is based on the adoption of the GME
formally defined by 
\begin{equation}
\frac{d}{dt}\mathbf{p}(t)=-\int_{0}^{t}\Phi (t-t^{\prime }{)\mathbf{Kp}}%
(t^{\prime })dt^{\prime },  \label{kenkre}
\end{equation}
where $\mathbf{p}\left( t\right) $ is the $m$-dimensional population vector
of $m$ sites, so that its $i$th component, $p_i(t)$, represents the probability of finding the walker at time $t$ in the $i$th site.
%for the flactuating variable $\xi$ to  be at time $t$ in the $i$th of its $m$ states.
$\mathbf{K}$ is a transition matrix between the sites and $%
\Phi (t)$ is the memory kernel. We study the case where the fluctuating
variable $\xi $ leading to this GME is dichotomous. This means that the GME
has only two states and the matrix $\mathbf{K}$ reads

\begin{equation}
\mathbf{K}=\left( 
\begin{array}{rr}
1 & -1 \\ 
-1 & 1
\end{array}
\right) .  \label{ctrw6}
\end{equation}
The waiting-time distribution in either of the two states is denoted by 
%\begin{equation}
%\label{experimental}
%\psi(t) = (\mu -1) \frac{T^{\mu -1}}{(t+T)^{\mu}}.
%\end{equation}
$\psi (t)$. We establish the connection of the GME with the CTRW \cite
{gerardo,montroll} by relating the Laplace transform of the memory kernel $%
\Phi (t)$, given by $\hat{\Phi}(u),$ to the Laplace transform of the
waiting-time distribution $\psi (t)$, given by $\hat{\psi}(u)$, as follows 
\begin{equation}
\hat{\Phi}(u)=\frac{u\hat{\psi}(u)}{1-\hat{\psi}(u)}.  \label{brandnew}
\end{equation}

We consider two distinct processes, that of preparation and that of
observation. The preparation process establishes the initial conditions of
the set of trajectories under study. A trajectory is a sequence of symbols $+
$ or $-$, specifying whether the system is in the state $|+\rangle $ or $%
|-\rangle $. We call the time interval with the system either entirely in
the state $|+\rangle $ or entirely in the state $|-\rangle ,$ the laminar
region. With Zumofen and Klafter \cite{klafter}, we assume that the preparation process, beginning at time $t=-t_{a}$%
, insures that all these trajectories begin with the system at the onset of
a laminar region, either $+$ or $-$.

The observation process begins at time $t=0\geq -t_{a}$. The distribution of
first sojourn times is denoted by $\psi _{t_{a}}(t)$ and is, in general,
different from $\psi (t)$. In fact, the laminar region corresponding to the
first sojourn time might have begun earlier than $t=0$. The exact expression
for this time distribution is 
\begin{equation}
\psi _{t_{a}}(t)=\psi (t+t_{a})+\sum_{n=1}^{\infty }\int_{0}^{t_{a}}dy\psi
(y+t)\psi _{n}(t_{a}-y),  \label{exactaging}
\end{equation}
where $\psi _{n}(\tau )$ denotes the probability that $n$ jumps occur during
the time interval of length $\tau $, the last of which occurs at time $%
t=\tau $. As is well known, for a renewal process the waiting times for
successively more jumps is given by the convolution 
\begin{equation}
\psi _{n}(t)=\int_{0}^{t}\psi _{n-1}(t^{\prime })\psi _{1}(t-t^{\prime})dt^{\prime },  \label{nasafunctionofn-1}
\end{equation}
with $\psi (t)\equiv \psi _{1}(t)$. The $t_{a}$-old distribution of first sojourn time was discussed earlier in some detail \cite{rejuvenation}. However, a careful analysis of Eq. (5) can help the reader to realize the rationale behind this distribution of sojourn times. %The age-dependent distribution of the first
%sojourn time was discussed earlier in some detail \cite{rejuvenation}.
%However, the reader can easily realize the rationale behind the age-dependent
%distribution of sojourn times.
The first term on the right hand side of Eq. (\ref{exactaging}) corresponds to the case when the first laminar region is
extended in time more than the time interval $t_{a}$ between preparation and
observation. The second term takes into account the cases when the last
laminar region began after the preparation time, after a sequence of an
arbitrarily large number of earlier laminar regions, the first of which, of
course, begins at $t=-t_{a}$.

In the Poisson case, because of its unique functional form, there is no
dependence of $\psi _{t_{a}}$ on $t_{a}$, and consequently no aging. In the
non-Poisson case, on the contrary, the two waiting-time distributions, $\psi
_{t_{a}}$ and $\psi (t)$, are identical only if $t_{a}=0$. In this case both
CTRW and GME correspond to switching-on the observation process at the same
time as the preparation process, and the connection between the two pictures
is given by Eq. (\ref{brandnew}).

Allegrini et al. \cite{gerardo} proved that the GME is compatible with an infinitely
aged CTRW, provided that the memory kernel $\Phi (t)$ is made compatible
with an infinitely aged condition, characterized by a distribution of first
sojourn times, which is infinitely aged. In this case $t_{a}\rightarrow
\infty $ so that 
\begin{equation}
\hat{\Phi}_{\infty }(u)=\frac{u\hat{\psi}_{\infty }(u)}{1+\hat{\psi}(u)-2\hat{\psi}_{\infty }(u)},  \label{memorykernel}
\end{equation}
where 
\begin{equation}
\psi _{\infty }(t)=\frac{1}{\langle \tau \rangle }\int_{t}^{\infty
}dt^{\prime }\psi (t^{\prime }).  \label{infinitelyaged}
\end{equation}
It is straigthforward to extend the calculations of Ref. \cite{gerardo} to
the case of an arbitrarily $t_{a}$-old system, so that Eq.(\ref{memorykernel}) is replaced with 
\begin{equation}
\hat{\Phi}_{t_{a}}(u)=\frac{u\hat{\psi}_{t_{a}}(u)}{1+\hat{\psi}(u)-2\hat{\psi}_{t_{a}}(u)},  \label{arbitraryage}
\end{equation}
where $\hat{\psi}_{t_{a}}(u)$ is the Laplace transform of $\psi _{t_{a}}(t)$%
. % whose explicit expression is \cite{prl}
%\begin{equation}\label{laplacewait}
%\psi _{u_{a}}(t)=\frac{1}{1-\psi (u_{a})}e^{u_{a}t}\left[\psi(u_{a})-\int_{0}^{t}e^{-u_{a}y}\psi (y)dy\right]
%\hat{\psi}_{t_{a}}(u)=\frac{(1-\hat{\psi}(u))(1-e^{-ut_{a}})+ue^{ut_{a}}
%\int_{0}^{t_{a}}e^{-uy}\Psi (y)dy}{ug(t_{a})}.
%\end{equation}

Using the GME with the $t_{a}$-old memory kernel, we define the age-dependent
correlation function $\Phi _{\xi }^{(t_{a})}(t)$ through its Laplace
transform, as follows 
\begin{equation}
\hat{\Phi}_{\xi }^{(t_{a})}(u)=\frac{1}{u+2\hat{\Phi}_{t_{a}}(u)}.
\label{density}
\end{equation}
%If we  indicate with $p_i(t)$ the $i$th component of the probability vector, 
This prescription corresponds to setting 
\begin{equation}
\Phi _{\xi }^{(t_{a})}(t)=\frac{p_1(t)-p_2(t)}{p_1(0)-p_2(0)},
\label{agingonsager}
\end{equation}
%with $p_i(t)$ representing the $i$th component of the probability vector,
 as one can easily check by Laplace transforming both sides and using the GME
with the $t_{a}$-old memory kernel (see section II B). 
%and then calculating the  Laplace transform emploing the GME with the $t_{a}$-old memory kernel.
In other words, the correlation function of arbitrary age mirrors the extension of the
Onsager principle, which is usually limited to infinitely aged systems \cite
{gerardo}, to physical conditions of arbitrary age. It has to be pointed out
that at this stage there is no guarantee that the  Onsager principle of arbitrary age
holds true. However, in Section \ref{sectiondensity} we show that Eq. (\ref{density}) yields the exact result of Godr\`{e}che and Luck \cite{luck} for
the corresponding correlation function. Furthermore, in Section \ref
{sectiontrajectory} we establish the same result using arguments based on
trajectories rather than densities, thereby affording an independent
construction of the exact expression for the  $t_{a}$-old correlation function. All
this can be thought of as a compelling demonstration of the correctness of of Eq. (\ref{agingonsager}), extending the Onsager principle to conditions of arbitrary age.
%the Onsager principle of arbitrary age of Eq. (\ref{agingonsager}).

\subsection{Derivation of the exact expression proposed by Godr\`{e}che and Luck: the probability perspective}

\label{sectiondensity}

To establish that the proposed approach yields the exact expression of Godr\`{e}che and Luck \cite{luck}, let us express the  $t_{a}$-old correlation
function through the probability vector $\mathbf{p}\equiv (p_{1},p_{2})$ for
the dichotomous variable $\xi =\pm 1$ to have either positive (state 1) or
negative (state 2) values. We have that: 
\begin{eqnarray}\label{correlaging} 
&&\Phi _{\xi }^{(t_{a})}(t)=\frac{\langle \xi (0)\xi (t)\rangle _{t_{a}}}{\langle \xi ^{2}\rangle }=\\
&&\nonumber p_{1}(0)p_{1}(t|1,t=0) +p_{2}(0)p_{2}(t|2,t=0)\\
&&\nonumber -p_{1}(0)p_{2}(t|1,t=0)-p_{2}(0)p_{1}(t|2,t=0), 
\end{eqnarray}
where $p_{j}(t|k,t=0)$ is the conditional probability that the variable $\xi 
$ is in the state $j$ at time $t$, given that at time $t=0$ it was in the
state $k$. This means that $p_{j}(t|k,t=0)$ is obtained letting those
trajectories evolve that at time $t=0$ had $\xi $ in the state $k$. For a
straightforward evaluation of $p_{j}(t|k,t=0)$, we use the GME formalism,
adapted to the $t_{a}$-old system, and we take into account the initial
condition %It straightforward to evaluate
%$p_{j}(t|k,t=0)$, we use the GME formalism, taking into account that
$p_{i}(0|k,t=0)=\delta _{i,k}$. %and the age $t_{a}$ of the process.
According to the GME the components of the conditional probability vector are determined
 %(adopting the convention of the sum  over repeated indexes), 
by:
\begin{equation}
\frac{d}{dt}p_{j}(t|k,t=0)=-\int_{0}^{t}dt^{\prime }\Phi
_{t_{a}}(t-t^{\prime })\sum_{i=1}^{2}\mathit{K}_{ji}p_{i}(t^{\prime }|k,t=0),
\label{GME}
\end{equation}
with $\hat{\Phi}_{t_{a}}(u)$ given by Eq. (\ref{arbitraryage}) and the elements of $%
\mathbf{K}$ given by Eq. (\ref{ctrw6}).

By Laplace transforming (\ref{GME}) and doing some algebra, we obtain 
\begin{equation}
\hat{p}_{j}(u|k,0)=\sum_{i=1}^{2}(u\mathbf{I}+\hat{\Phi}_{t_{a}}(u)\mathbf{K}%
)_{ji}^{-1}p_{i}(0|k,0).
\end{equation}
Defining the matrix 
\begin{equation}
\mathbf{J}=(u\mathbf{I}+\hat{\Phi}_{t_{a}}(u)\mathbf{K})^{-1}=\left( 
\begin{array}{cc}
\frac{u+\hat{\Phi}_{t_{a}}(u)}{u\left[ u+2\hat{\Phi}_{t_{a}}(u)\right] } & 
\frac{\hat{\Phi}_{t_{a}}(u)}{u\left[ u+2\hat{\Phi}_{t_{a}}(u)\right] } \\ 
\nonumber \frac{\hat{\Phi}_{t_{a}}(u)}{u\left[ u+2\hat{\Phi}%
_{t_{a}}(u)\right] } & \frac{u+\hat{\Phi}_{t_{a}}(u)}{u\left[ u+2\hat{\Phi}%
_{t_{a}}(u)\right] }
\end{array}
\right) ,  \label{inverse}
\end{equation}
and using the initial condition $p_{i}(0|k,t=0)=\delta _{i,k}$, we obtain
for the Laplace transform of the conditional probability vector the
following expression 
\begin{equation}\label{pj}
\hat{p}_{j}(u|k,0)=\frac{\left[ u+\hat{\Phi}_{t_{a}}(u)\right]\delta _{j,k}+\hat{\Phi}_{t_{a}}(u)\left(\delta _{j,k+1}+\delta _{j,k-1}\right)}{u\left[ u+2\hat{\Phi}_{t_{a}}(u)\right] }.  
\end{equation}
%To simplify (\ref{pj}) we<Laplace trasform (\ref{correlaging}) and use (\ref{pj}) 
Using Eq. (\ref{pj}) we can Laplace trasform (\ref{correlaging}) to obtain 
%\begin{eqnarray}
%\nonumber \hat{\Phi}_{\xi }^{(t_{a})}(u) &=&p_{1}(0)\frac{u+\hat{\Phi}_{t_{a}}(u)}{u( u+2\hat{\Phi}_{t_{a}}(u)) %}-p_{1}(0)\frac{\hat{\Phi}_{t_{a}}(u)}{u(u+2\hat{\Phi}_{t_{a}}(u))}\\
%&+& p_{2}(0)\left[\frac{u+\hat{\Phi}_{t_{a}}(u)}{u\left[ u+2\hat{\Phi}_{t_{a}}(u)\right] }-\frac{\hat{\Phi}_{t_{a}}(u)}{u\left[ u+2\hat{\Phi}_{t_{a}}(u)\right] }\right]
%%p_{1}(0)\frac{u+\hat{\Phi}_{t_{a}}(u)}{u\left[ u+2\hat{\Phi}_{t_{a}}(u)\right] }+p_{2}(0)\frac{u+\hat{\Phi}_{t_{a}}(u)}{u\left[ u+2\hat{\Phi}_{t_{a}}(u)\right] }  \nonumber \\
%%&-&p_{1}(0)\frac{\hat{\Phi}_{t_{a}}(u)}{u\left[ u+2\hat{\Phi}_{t_{a}}(u)\right] }-p_{2}(0)\frac{\hat{\Phi}_{t_{a}}(u)}{u\left[ u+2\hat{\Phi}_{t_{a}}(u)\right] }.
%\end{eqnarray}
\begin{equation}
\begin{split}
\hat{\Phi}_{\xi }^{(t_{a})}(u)& =p_{1}(0)\frac{u+\hat{\Phi}_{t_{a}}(u)}{u\left( u+2\hat{\Phi}_{t_{a}}(u)\right) }+p_{2}(0)\frac{u+\hat{\Phi}_{t_{a}}(u)}{u\left(u+2\hat{\Phi}_{t_{a}}(u)\right)}\\
&- p_{1}(0)\frac{\hat{\Phi}_{t_{a}}(u)}{u\left( u+2\hat{\Phi}_{t_{a}}(u)\right)}-p_{2}(0)\frac{\hat{\Phi}_{t_{a}}(u)}{u\left(u+2\hat{\Phi}_{t_{a}}(u)\right) }
%p_{1}(0)\frac{u+\hat{\Phi}_{t_{a}}(u)}{u\left[ u+2\hat{\Phi}_{t_{a}}(u)\right] }+p_{2}(0)\frac{u+\hat{\Phi}_{t_{a}}(u)}{u\left[ u+2\hat{\Phi}_{t_{a}}(u)\right] }  \nonumber \\
%&-p_{1}(0)\frac{\hat{\Phi}_{t_{a}}(u)}{u\left[ u+2\hat{\Phi}_{t_{a}}(u)\right] }-p_{2}(0)\frac{\hat{\Phi}_{t_{a}}(u)}{u\left[ u+2\hat{\Phi}_{t_{a}}(u)\right] }.
\end{split}
\end{equation}
We note that the probability is normalized, $p_{1}(0)+p_{2}(0)=1$. Thus, it
follows that 
%\begin{eqnarray}\label{result1}
%\hat{\Phi}_{\xi }^{(t_{a})}(u)&=&\left[
%\frac{u+\hat{\Phi}_{t_{a}}(u)}{u\left(u+2\hat{\Phi}_{t_{a}}(u)\right)}-\frac{\hat{\Phi}_{t_{a}}(u)}{u\left(u+2\hat{\Phi}_{t_{a}}(u)\right)}\right] \left[ %p_{1}(0)+p_{2}(0)\right]\\
%&&\nonumber  =\frac{1}{u+2\hat{\Phi}_{t_{a}}(u)},  
%\end{eqnarray}
\begin{equation}\label{result1}
\begin{split}
\hat{\Phi}_{\xi }^{(t_{a})}(u)&=\left[ \frac{u+\hat{\Phi}_{t_{a}}(u)}{u\left(u+2\hat{\Phi}_{t_{a}}(u)\right)}-\frac{\hat{\Phi}_{t_{a}}(u)}{u\left(u+2\hat{\Phi}_{t_{a}}(u)\right)}\right] \left[ p_{1}(0)\right.\\
&\left. +p_{2}(0)\right]=\frac{1}{u+2\hat{\Phi}_{t_{a}}(u)},  
\end{split}
\end{equation}
confirming the correctness of the definition introduced in  Eq. (\ref{density}).
Substituting (\ref{arbitraryage}) into (\ref{result1}) we obtain 
\begin{equation}
\hat{\Phi}_{\xi }^{(t_{a})}(u)=\frac{1}{u\left[ 1+2\frac{\hat{\psi}%
_{t_{a}}(u)}{1+\hat{\psi}(u)-2\hat{\psi}_{t_{a}}(u)}\right] }=\frac{1}{u}\left[ 1-\frac{2\hat{\psi}_{t_{a}}(u)}{1+\hat{\psi}(u)}\right] ,
\label{exact1}
\end{equation}
which coincides with the results of Godr\`{e}che and Luck \cite{luck}.
Furthermore, the ratio of the differences in probability is determined in
Laplace space by 
\begin{eqnarray} \label{exactprobability}
\frac{\hat{p}_{1}(u)-\hat{p}_{2}(u)}{p_{1}(0)-p_{2}(0)}&=&\sum_{i=1}^{2}\frac{(J_{1i}-J_{2i})p_{i}(0)}{p_{1}(0)-p_{2}(0)}\\
&=&\nonumber \frac{1}{u+2\hat{\Phi}_{t_{a}}(u)}=\hat{\Phi}_{\xi }^{(t_{a})}(u). 
\end{eqnarray}
As pointed out earlier, Eq.(\ref{exactprobability}) means that one can
extend the Onsager Principle from the infinitely aged systems, for which Onsager
originally defined it, to  systems of any age. In the latter case the relaxation is proportional to the $t_{a}$-old correlation function, not to the infinitely old, or equilibrium, correlation function. In summary, we discovered an Onsager principle of arbitrary age, at least  in the special case of the dichotomous variables considered in this paper.
%The relaxation in the latter system
%is proportional to the correlation function of arbitrary age, not to the unperturbed
%correlation function. In summary, we discover an ``age-dependent'' Onsager
%regression principle, at least in the special case of the dichotomous
%variables considered in this paper.

\subsection{Derivation of the exact expression proposed by Godr\`{e}che and
Luck: the trajectory perspective}

\label{sectiontrajectory}

It is possible to again derive the exact result of Eq. (\ref{exact1}) from a
different perspective, which will allow us, in Section \ref{trajpersp}, to
propose an analytic expression for the $t_{a}$-old correlation function as a
function of time. This expression, as we shall see, is not exact, but it is
shown numerically to be a very good approximation to the exact result.

The usual method of connecting the correlation function $\Phi _{\xi }$ to
the waiting-time distribution, within a trajectory perspective, is to
introduce a theoretical waiting-time distribution, $\psi ^{*}(t)$, which
cannot be observed directly. In fact, the experimental waiting-time
distribution, namely, the distribution of times with alternate signs,
denoted by us as $\psi (t)$, is obtained from the theoretical waiting-time
distribution, $\psi ^{*}(t)$, by adopting the following procedure. We divide
the time axis into bins, whose size is determined by the waiting-time
distribution $\psi^*(t)$. Then, these bins are assigned either the value $1$
or the value $-1$, by tossing a coin to make the decision. It is evident
that the intervals along the time axis with the same sign, are larger than
the time bins determined by $\psi^*(t)$, since two or more consecutive coin
tossings might have produced the same sign. It is shown \cite{klafter} that
the Laplace transform of $\psi $ is connected to the Laplace transform of $%
\psi ^{*}$ via the relation 
\begin{equation}
\hat{\psi}(u)=\frac{\hat{\psi}^{*}(u)}{2-\hat{\psi}^{*}(u)}.
\label{usualpsistar}
\end{equation}
Let us use the term \emph{event} to denote the coin tossing introduced
above. The expression (\ref{usualpsistar}) is the result of summing over all
possibilities of not changing sign with a coin toss, which turns out to be a
geometrical series in the Laplace representation. The correlation function $%
\Phi _{\xi }$ and the theoretical waiting-time distribution function $\psi
^{*}$ are connected through the relation 
\begin{equation}
\Phi _{\xi }(t)=\frac{\int_{t}^{\infty }(t-\tau )\psi ^{*}(\tau )d\tau }{%
\langle \tau \rangle },  \label{window}
\end{equation}
where the average waiting time is given by 
\begin{equation}
\langle \tau \rangle \equiv \int_{0}^{\infty }\tau \psi ^{*}(\tau )d\tau .
\end{equation}
Eq. (\ref{window}) determines that the correlation function $\Phi _{\xi }(t)$
is equal to the probability of finding a window of length $t$ without
internal events.

The same result can be immediately recovered using 
\begin{eqnarray}\label{alternative}
\Phi _{\xi }(t)&=&\Psi _{\infty }^{*}(t)\equiv \int_{t}^{\infty }dt^{\prime}\psi _{\infty }^{*}(t^{\prime })\\
&=&\nonumber \frac{1}{\langle\tau \rangle}\int_{t}^{\infty }dt^{\prime }\int_{t^{\prime }}^{\infty }dt^{\prime\prime}\psi ^{*}(t^{\prime \prime}).  
\end{eqnarray}
We note that, see Ref. \cite{gerardo}, 
\begin{equation}
\psi _{\infty }^{*}(t)\equiv \frac{1}{\langle\tau \rangle}\int_{t}^{\infty
}dt^{\prime }\psi ^{*}(t^{\prime })
\end{equation}
is the infinitely aged waiting-time distribution. Actually, Eq. (\ref
{alternative}) is the infinitely aged correlation function, a special case
of the more general prescription 
\begin{equation}
\Phi _{\xi }^{(t_{a})}(t)=\Psi _{t_{a}}^{*}(t)\equiv \int_{t}^{\infty}dt^{\prime }\psi _{t_{a}}^{*}(t^{\prime })=1-\int_{0}^{t}dt^{\prime }\psi_{t_{a}}^{*}(t^{\prime }).  \label{alternativeofanyage}
\end{equation}

It is straightforward to show that the $t_{a}$-old experimental waiting-time
distribution and the $t_{a}$-old theoretical waiting-time distribution are
connected through the following sum of convolutions: 
%It is straightforward to show that the age-dependent  waiting-time distribution
%function is the sum of convolutions
\begin{eqnarray} \label{convol}
\psi _{t_{a}}(t)&=&\frac{1}{2}\psi _{t_{a}}^{*}(t)\star \left\{ \delta (t)+\frac{1}{2}\psi ^{*}(t)\right.\\
&& \nonumber +\left.\frac{1}{2}\psi ^{*}(t)\star \frac{1}{2}\psi^{*}(t)+\cdots \right\} , 
\end{eqnarray}
where the symbol $\star $ denotes time convolution. 
In fact, after a first interval of time followed by a coin toss with no
change of sign, determined by $\psi _{t_{a}}^{*}/2$, the next intervals of
time with no change of sign according to the coin tossing prescription, are
determined by $\psi ^{*}/2$. The sum of the convolutions takes into account
all the possible sequences of intervals of time with no change of sign
before a change of sign of the variable $\xi $ eventually occurs, and gives as 
final result the distribution for a first observed sojourn time $t$ of the
variable $\xi $ in one of its two states, that is $\psi _{t_{a}}(t)$.

%In fact, after the first aged sojourn, with the factor $1/2$, dictated by the coin tossing
%prescription, the next sojourn times  with another factor of $1/2$, are
%determined by the waiting-time distribution $\psi ^{*}(t)$.
Thus by summing the geometric series in the Laplace variables, from (\ref
{convol}) we obtain 
\begin{equation}
\hat{\psi}_{t_{a}}(u)=\frac{\hat{\psi}_{t_{a}}^{*}(u)}{2-\hat{\psi}^{*}(u)}.
\label{gerardomifafaticare1}
\end{equation}
Using Eq. (\ref{usualpsistar}) we write Eq. (\ref{gerardomifafaticare1}) as 
\begin{equation}
\hat{\psi}_{t_{a}}^{*}(u)=\frac{2\hat{\psi}_{t_{a}}(u)}{1+\hat{\psi}(u)}.
\label{gerardomifafaticare2}
\end{equation}
%Let us Laplace transforming  Eq. (\ref{alternativeofanyage}) and use Eq. (\ref
%{gerardomifafaticare2}), to obtain
By Laplace transforming Eq. (\ref{alternativeofanyage}) and using Eq. (\ref
{gerardomifafaticare2}), we obtain 
\begin{equation}
\hat{\Phi}_{\xi }^{(t_{a})}(u)=\frac{1-\hat{\psi}_{t_{a}}^{*}(u)}{u}=\frac{1%
}{u}\left[ 1-\frac{2\hat{\psi}_{t_{a}}(u)}{1+\hat{\psi}(u)}\right] ,
\label{exact2}
\end{equation}
namely, we again recover the exact result of Godr\`{e}che and Luck given by
Eq. (\ref{exact1}). This establishes the equivalence of the trajectory and
GME prescriptions for this process.

\subsection{Generalized Master Equation of arbitrary age}

We are now in a position to make a preliminary balance of the results
obtained so far. The first is that we have generalized the result of an
earlier paper \cite{gerardo}, in that we have derived the GME of arbitrary age

\begin{equation}
\frac{d}{dt}\mathbf{p}(t)=-\int_{0}^{t}\Phi _{t_{a}}(t-t^{\prime }{)\mathbf{%
Kp}}(t^{\prime })dt^{\prime },  \label{aginggme}
\end{equation}
whose memory kernel $\Phi _{t_{a}}(t)$ is defined though its Laplace
transform by means of Eq. (\ref{arbitraryage}). We have also shown that the
Onsager principle, valid for infinitely aged systems, can be extended to
conditions of any age, and that this extension allows us to derive an exact
expression for the $t_{a}$-old correlation function. However, the analytic results
obtained so far are in the Laplace domain. It is desireable to achieve them
in the time domain, as well. We now turn our attention to the latter.

\section{The Liouville-like approach}

\label{liouvappr}

In this section we derive another expression for the $t_{a}$-old correlation
function  adopting a perspective where aging is determined by the out of
equilibrium bath for the variable of interest, $\xi $. There might exist
conditions, as we shall see, where equilibrium is not even allowed. The
expression for the $t_{a}$-old correlation function  afforded by this perspective
is exact, and is thus equivalent to the Godr\`{e}che and Luck expression of
Eq. (\ref{exact1}). However, the exact expression is implicit, rather than
explicit, and is therefore more convenient for the numerical calculations
done subsequently.

Here we adopt the perspective of earlier work \cite{chemphys, paper2} to
account for the aging effects characterizing the fluctuations of the
dichotomous variable $\xi $. These fluctuations occur while the environment
of the variable, $\xi $, slowly drifts. This drifting process is extended
over time, and could lead to circumstances where it is not possible to
attain equilibrium asymptotically. In keeping with the jargon of statistical
mechanics the environmental or ``irrelevant'' variable is called $y$ and in
the model moves in the interval $I=[0,2]$. In the semi-interval $[0,1]$, we
use the equation of motion for the probability density
\begin{equation}
\frac{\partial p(y,t)}{\partial t}=-\lambda \frac{\partial }{\partial y}%
y^{z}p(y,t),  \label{cheerfulequationwithnobackinjection}
\end{equation}
This is the motion determined by a potential, with the minimum at $y=1$, in
the over-damped case. %The parameter $z$ is defined by
%\begin{equation}
%z \equiv \frac{\mu}{\mu-1}.
%\end{equation}
In the interval $[1,2]$, the over-damped potential is the mirror image of
the potential acting on the left interval. Consequently, if the initial
condition is located in the internal part of the interval $[0,1]$, the
particle moves, from the left to the right, with a deterministic motion,
towards $y=1$. If the particle is initially located in the interior of the
interval $[1,2]$, it moves deterministically from the right to the left.
When the particle reaches the potential minimum, it is injected back, with
equal probability, into any of the points of the interval $I$ excluding $y=0$%
, $y=1$ and $y=2$. The time spent by $y$ within $I$ corresponds to
sojourning in one of the two states of the variable $\xi $, either $%
|+\rangle $ or $|-\rangle $. The instant of back injection corresponds to
the choice of the new state and, with equal probability, this is either the
same state or the other state. The variable $y$ represents the environment
of the variable $\xi $, and its initial distribution is given by the state
of the bath. The corresponding waiting-time distribution between two
consecutive back injections is 
\begin{equation}
\psi ^{*}(t)=(\mu -1)\frac{T^{\mu -1}}{(t+T)^{\mu }},  \label{psistarT}
\end{equation}
%where the parameter $z$ of the dynamical model is chosen via the formula 
where the index $\mu $ is related to $z$ of Eq. (\ref
{cheerfulequationwithnobackinjection}) by 
\begin{equation}
\mu \equiv \frac{z}{z-1},
\end{equation}
and the parameter $T$, characterizing the waiting-time distribution of Eq. (%
\ref{psistarT}) is %in the waiting-time distribution is determined by
%normalization to be 
\begin{equation}
T\equiv \frac{\mu -1}{\lambda },
\end{equation}
in accordance with the normalization constraint. The authors of Refs. \cite
{gerardo,paper1} have shown that the essential properties of the dichotomous
non-Poisson fluctuation can be accounted for by limiting ourselves to this
simplified picture, involving only the semi-interval $[0,1]$. In this simple
picture the aging process is described by 
\begin{equation}
\frac{\partial p(y,t)}{\partial t}=-\lambda \frac{\partial }{\partial y}y^{z}p(y,t)+\lambda p(1,t),  \label{cheerfulequation}
\end{equation}
which takes into account the back injection into the semi-interval,
occurring with uniform probability, when the particle reaches the point $y=1$%
. For simplicity, but with no loss of generality, we fix $\lambda =1$.

Using the results of Ref. \cite{paper2} the $t_{a}$-old correlation function is
evaluated as follows. The bath is prepared at time $t=-t_{a}$. This means
that at time $t=-t_{a}$ the distribution of $y$ within the interval $I$ is
flat. In the case $z<2$ this distribution tends to the equilibrium
distribution, $p_{eq}\propto \frac{1}{y^{z-1}}$. If $z>2$ this distribution
diverges, thereby implying that the distribution approaches the Dirac delta
function located at $y=0$. This is the non-stationary condition, the
condition where equilibrium is not allowed, and only a condition of eternal
drift is admitted. Suppose this distribution evolve for a time $t_{a}$,
without our observing it. This means that we begin the observation process
when the system has a new distribution, different from the initial flat
distribution, and determined by its time evolution from $t=-t_{a}$ to $t=0$,
described by Eq. (\ref{cheerfulequation}). 
%This failure to observe the system
%means that the process of observation begins at $t=0$, implying that
%Eq. (\ref{cheerfulequation}) is used to determine the distribution at ime $t=0$.
For $t\geq 0$ Eq. (\ref{cheerfulequation}) is replaced with Eq. (\ref
{cheerfulequationwithnobackinjection}) 
\begin{equation}
\frac{\partial p(y,t)}{\partial t}=-\frac{\partial }{\partial y}y^{z}p(y,t),
\label{observation}
\end{equation}
namely the back injection process is stopped, thereby implying that the
population decreases.

The theory of Ref. \cite{paper2} relates the probability solution to the
Liouville-like equation to the $t_{a}$-old correlation function as follows
\begin{equation}
\Phi _{\xi }^{(t_{a})}(t)=\int_{0}^{1}dyp(y,t).  \label{angeloisagenius}
\end{equation}
Note that the initial condition $p(y,0)$ is obtained from Eq. (\ref
{cheerfulequation}) moving from the flat distribution at $t=-t_{a}$. The
time evolution, corresponding to the observation process, is determined by
Eq. (\ref{observation}). This simple picture is a fair representation of the
description made in terms of trajectories in Section \ref{redpersp}. The
fact that the norm of $p(y,t)$, when $p(y,t)$ is described by Eq. (\ref
{observation}), is not conserved, reflects the occurrence of jumps from one
state to the other, making the population of a given state decrease. Let us
remark that the correlation function is determined by the antisymmetric part
of the whole distribution \cite{paper2}. This formal condition corresponds
to observing the time evolution of $p(y,t)$, with $y$ ranging from $y=0$ to $%
y=1$, under the action of Eq.(\ref{cheerfulequationwithnobackinjection}),
with no back injection process. The process of back injection is essential
to determine the correlation function of arbitrary age. However, once the parameter $%
t_{a}$ is fixed, and with it the age of the correlation function, the
evaluation of the correlation function is done by imagining the bath frozen
in the distribution corresponding to this fixed age. This is the reason we
use Eq. (\ref{cheerfulequationwithnobackinjection}) to determine the
correlation function rather than Eq. (\ref{cheerfulequation}).

The process of back injection is essential for the slow environmental drift,
but the correlation time of age $t_{a}$ is determined by the age-fixed
condition, the bath being $t_{a}$-old, and keeping this age forever.

Our goal is to calculate the correlation function of arbitrary age. According to Eq. (%
\ref{angeloisagenius}), this goal requires that we evaluate $p(y,t)$, first.
To carry out this calculation we follow the procedure illustrated in detail
in Ref. \cite{massi}. The time evolution from $t=-t_{a}$ to $t<0$ is
described by Eq. (\ref{cheerfulequation}), yielding for the solution
\begin{eqnarray}
p(y,t)&=&\frac{1}{[1+(z-1)(t+t_{a})y^{z-1}]^{z/(z-1)}}\\
&+&\nonumber\int_{-t_{a}}^{t}\frac{p(1,\tau )d\tau }{[1+(z-1)(t-\tau )y^{z-1}]^{z/(z-1)}},
\end{eqnarray}
with $p(1,\tau )$ determined, in turn, by Eq. (\ref{cheerfulequation}).

As already stated, for $t>0,$ $p(y,t)$ is governed by Eq. (\ref{observation}%
), so that its solution is, given the initial value,
\begin{equation}
p(y,t)=\frac{ p(\eta,0)}{[1+(z-1)y^{z-1}t]^{z/(z-1)}}|_{\eta =\frac{y}{[1+(z-1)y^{z-1}t]^{1/(z-1)}}}
\end{equation}
Now, we can calculate the final expression for the age-dependent correlation
function  given by Eq. (\ref{angeloisagenius}). After some simple algebra
we obtain the following expression: 
\begin{eqnarray}\label{newsol}
\Phi _{\xi }^{(t_{a})}(t)&=&\frac{1}{[1+(z-1)(t+t_{a})]^{1/(z-1)}}\\
&+&\nonumber \int_{-t_{a}}^{0}\frac{p(1,\tau )d\tau }{[1+(z-1)(t-\tau )]^{1/(z-1)}} \\
&=&\nonumber \Psi ^{*}(t+t_{a})+\int_{-t_{a}}^{0}\frac{p(1,\tau )d\tau }{[1+(z-1)(t-\tau )]^{1/(z-1)}} \\
&=&\nonumber \Psi ^{*}(t+t_{a})+[1-\Psi ^{*}(t_{a})]\\
&+&\nonumber \int_{-t_{a}}^{0}\frac{p(1,\tau)d\tau }{[1+(z-1)(t-\tau )]^{1/(z-1)}}\\
&-&\nonumber\int_{-t_{a}}^{0}\frac{p(1,\tau )d\tau }{[1-(z-1)\tau ]^{1/(z-1)}}.
\end{eqnarray}
Note that in Eq.(\ref{newsol}) we are using the definition 
\begin{equation}
\Psi ^{*}(t)=\int_{t}^{\infty }\psi ^{*}(\tau )d\tau .  \label{survivalstar}
\end{equation}
This notation for the survival probability is also used in Section \ref
{paoloa}. We notice that, $t>0$ and $p(1,t)$ is calculated from Eq. (\ref
{cheerfulequation}) with $y=1$ and $-t_{a}<t<0$. This is an exact expression
that turns out to be convenient to numerically check the prescriptions of
Section \ref{trajpersp}.

It is worth using the perspective of this section to shed further light into
the problem under discussion in this paper. Let us notice that

\begin{itemize}
\item  if the age vanishes $t_{a}\rightarrow 0$, then the $t_{a}$-old correlation
function reduces to the probability of no event occuring $\Phi _{\xi
}^{(t_{a})}(t)\rightarrow \Psi ^{*}(t)$.

\item  if the age increases without limit $t_{a}\rightarrow \infty $, then
the $t_{a}$-old correlation function  reduces to a known result $\Phi _{\xi
}^{(t_{a})}(t)\rightarrow \Phi _{\xi }(t)$, where 
\begin{equation}
\Phi _{\xi }(t)=\frac{1}{[1+(z-1)t]^{(2-z)/(z-1)}}  \label{eqcorrfunct}
\end{equation}
\end{itemize}
is the usual correlation function calculated at equilibrium. So, we must
observe a crossover between two distinct regimes. Moreover, we anticipate
that we recover these results from a different perspective in Sec. \ref
{trajpersp}.

The perspective adopted in this section makes it easier to understand, on
physical grounds, why the Poisson condition produces aging. The Poisson
condition corresponds to $z=1$ in the nonlinear dynamical map, and in this
case the distribution is always flat. Aging is the process of slow
regression of the bath variable to equilibrium, a process that can, in
principle, last forever. With this representation it is possible to study
the correlation function of arbitrary age also in the case $\mu <2$. Allegrini \emph{et
al.} \cite{paper2} give a detailed discussion of the effect of external
perturbations.

In the case $z<2$ equilibrium is possible. An external perturbation can be
used to create a condition whereby the system is out of equilbrium. The
observation of the population difference between the two states, $\xi =1$
and $\xi =-1$, is equivalent to determining the correlation function. Note
that the initial condition that we use to study the time evolution of $%
p(y,t) $ during the observation process, Eq. (\ref{observation}), is the
antisymmetric distribution of the interval $[0,2]$ folded into $[0,1]$ (see Ref. \cite
{paper2}). It is evident that in practice the equilibrium correlation
function is never observed. In fact, the external perturbation should create
a non-symmetric initial condition in the whole interval $[0,2]$. The anti-symmetric portion of this distribution, folded into the interval [0,1], should create the same out of equilibrium distribution as that produced by folding in the same way the anti-symmetric portion of the equilibrium distribution.
%\cite{paper2}.
 As pointed out in  Ref. \cite{paper2}, this is impossible to do in practice with a a perturbation lasting for a finite time, and, thus, with any realistic experimental procedure.
%The
%antisymmetric portion of this distribution, folded into the interval $I$
%should create the equilibrium distribution. This is impossible to do in
%practice with a perturbation lasting for a finite length of time.

\section{An analytical expression of the correlation function of arbitrary age}

\label{trajpersp} The exact expressions, Eq. (\ref{exact1}) and Eq. (\ref
{exact2}), allow us to determine the dependence of the age-dependent correlation
function on time via anti-Laplace transforms. This inversion is done
numerically since we are unable to carry out direct analytic inversion of
the Laplace expression. There is therefore the need to find an analytical
expression for $\Phi _{\xi }^{\left( t_{a}\right) }(t)$, even if the price
we pay is an approximation and the loss of exactness. This section is
devoted to the derivation of an accurate, albeit approximate, analytical
expression for the correlation function of arbitrary age.
\subsection{An exact implicit expression for the correlation function of arbitrary age}
\label{paoloa}
We write $\Phi _{\xi }^{(t_{a})}(t)$, namely the probability of not finding
events inside the interval $(0,t)$, given the occurance of an event at time $%
-t_{a}$, in the following implicit form 
\begin{equation}
\Phi _{\xi }^{(t_{a})}(t)=\Psi ^{*}(t+t_{a})+\int_{0}^{t_{a}}dt_{l}\psi
^{*}(t_{a}-t_{l})\Phi _{\xi }^{(t_{l})}(t).  \label{implicitform}
\end{equation}
This is the probability of finding an event at a time larger than $t$ from
the preceding one. The right-hand side of Eq. (\ref{implicitform}) is the
sum of two terms, the first corresponds to no event occurring in the whole
interval $(-t_{a},t)$ and the second term takes into account the possibility
of at least one event occurring between $-t_{a}$ and $0$. The instant $-t_{l}
$ signals the first of these events (or the event, if there is only one).
The coin-tossing procedure has the double effect of \emph{\ rejuvenating}
the process (the correlation function now has age $t_{l}$) and of factoring
the two probability functions inside the integral.

It is interesting to notice that Eq. (\ref{implicitform}) is exact, which
can be verified by differentiating both sides of the equation, and making
use of (\ref{alternativeofanyage}) and (\ref{survivalstar}). This procedure
yields
\begin{equation}
\psi _{t_{a}}^{*}(t)=\psi ^{*}(t+t_{a})+\int_{0}^{t_{a}}dt_{1}\psi
^{*}(t_{a}-t_{1})\psi _{t_{1}}^{*}(t),  \label{implicitagedpsi}
\end{equation}
namely an implicit (but analytic) expression for $\psi _{t_{a}}^{*}$. A
straighforward iteration of (\ref{implicitagedpsi}) leads to
%\begin{eqnarray}
%\psi_{t_{a}}^{*}(t) &=&\psi ^{*}(t+t_{a})+\int_{0}^{t_{a}}dt_{1}\psi
%^{*}(t_{a}-t_{1})\psi ^{*}(t+t_{1})  \label{iterate} \\
%&&+\int_{0}^{t_{a}}dt_{1}\int_{0}^{t_{1}}dt_{2}\psi ^{*}(t_{a}-t_{1})\psi
%^{*}(t_{1}-t_{2})\psi ^{*}(t+t_{2})  \nonumber \\
%&&+\int_{0}^{t_{a}}dt_{1}\int_{0}^{t_{1}}dt_{2}\int_{0}^{t_{2}}dt_{3}\psi
%^{*}(t_{a}-t_{1})\psi ^{*}(t_{1}-t_{2})\\
%&&\nonumber \psi ^{*}(t_{2}-t_{3})\psi
%^{*}(t+t_{3})  \nonumber +\cdots ,  
%%\psi^*(t+t_{a})+\sum_{n=1}^{\infty }\int_{0}^{t_{a}}dy\psi(y+t)\psi _{n}(t_{a}-y)
%\end{eqnarray}
\begin{equation}\label{iterate}
\begin{split}
\psi_{t_{a}}^{*}(t) &=\psi ^{*}(t+t_{a})+\int_{0}^{t_{a}}dt_{1}\psi^{*}(t_{a}-t_{1})\psi ^{*}(t+t_{1})   \\
&+\int_{0}^{t_{a}}dt_{1}\psi ^{*}(t_{a}-t_{1})\int_{0}^{t_{1}}dt_{2}\psi^{*}(t_{1}-t_{2})\psi ^{*}(t+t_{2})   \\
&+\int_{0}^{t_{a}}dt_{1}\psi^{*}(t_{a}-t_{1})\int_{0}^{t_{1}}dt_{2}\psi ^{*}(t_{1}-t_{2})\\
&\times\int_{0}^{t_{2}}dt_{3}\psi ^{*}(t_{2}-t_{3})\psi^{*}(t+t_{3})  +\cdots ,  
%\psi^*(t+t_{a})+\sum_{n=1}^{\infty }\int_{0}^{t_{a}}dy\psi(y+t)\psi _{n}(t_{a}-y)
\end{split}
\end{equation}
i.e., a sum over all possibilities of finding events at both times $t$ and $-t_{a}$, with no events between $0$ and $t$. 
This is exactly the definition of $\psi_{t_{a}}^{*}(t)$.
%, equivalent to Eq. (\ref{exactaging}) for $\psi_{t_{a}}(t)$. 
In detail, the first term in the right-hand side
of (\ref{iterate}), corresponds to the case in which no event occurs in the
interval $I\equiv (-t_{a},0]$, while all further terms refer, respectively,
to one event at $t_{1}$ in $I$, to two events, at $t_{1}$ and $t_{2}$, in $I$%
, and so on, with $t_{1}<t_{2}<t_{3}<\cdots $.
%This is exactly the definition of $\psi_{t_{a}}^{*}(t)$,
% and in fact regrouping the convolutions and with the correspondence $\psi(t) \to \psi^*(t)$ is equivalent to the definition of $\psi_{t_{a}}(t)$ given 
%by Eq. (\ref{exactaging}).
 Note that $\psi_{t_{a}}^{*}(t)$    is   obtained from $\psi^*(t)$ in the same way $\psi_{t_{a}}(t)$ is obtained from $\psi(t)$:  in fact if  in Eq. (\ref{exactaging}) we make the substitution $\psi(t) \to \psi^*(t)$ we obtain exactly Eq. (\ref{iterate}), i.e. the definition of $\psi_{t_{a}}^{*}(t)$.

\subsection{Equivalence with the exact expression of Godr\`{e}che and Luck}

As we said, Eq. (\ref{implicitform}) is exact. Consequently, this equation
is expected to be equivalent to the exact expression proposed by Godr\`{e}%
che and Luck in Ref. [7]. In this subsection, as a double check, we prove
that Eq. (\ref{implicitform}) is in fact equivalent to the expression of Godr%
\`{e}che and Luck. 
%In this sub{\bf section} we show that Eq. (\ref{implicitform}) is exact and
%equivalent to what obtained in \cite{luck}.
To do that, let us differentiate both sides of this equation with respect to 
$t$ to obtain 
\begin{equation}
\frac{d}{dt}\Phi _{\xi }^{(t_{a})}(t)=-\psi
^{*}(t+t_{a})+\int_{0}^{t_{a}}dt_{l}\psi ^{*}(t_{a}-t_{l})\frac{d}{dt}\Phi
_{\xi }^{(t_{l})}(t).  \label{diffimplicitform}
\end{equation}
A more tractable expression is obtained by taking the Laplace transform of
this new expression with respect to $t_{a}$ to obtain 
\begin{eqnarray}
\frac{d}{dt}\hat{\Phi}_{\xi }^{(s)}(t)&=&-e^{st}\left[ \hat{\psi ^{*}}%
(s)-\int_{0}^{t}dye^{-sy}\psi ^{*}(y)\right]\\
 &+&\nonumber\hat{\psi ^{*}}(s)\frac{d}{dt}\hat{\Phi}_{\xi }^{(s)}(t), \label{LTdiffimplicitform}
\end{eqnarray}
from which it follows that 
\begin{equation}
\frac{d}{dt}\hat{\Phi}_{\xi }^{(s)}(t)=-\frac{e^{st}\left[ \hat{\psi ^{*}}(s)-\int_{0}^{t}dye^{-sy}\psi ^{*}(y)\right] }{1-\hat{\psi ^{*}}(s)}.
\label{LTdiffimplicitform_bis}
\end{equation}
The right-hand side of this equation appears formidable, but
one can easily see by Laplace transforming Eq. (\ref{implicitagedpsi}) with respect to $t_a$,
%or from Eq. (15) of Ref. \cite{prl},  by identifying $\psi (t)$ with $\psi^{*}(t)$, 
that 
%(\ref{LTdiffimplicitform_bis}) is seen to be 
it is nothing more than the
Laplace transform with respect to $t_{a}$ of $-\psi _{t_{a}}^{*}(t)$, namely 
$-\hat{\psi}_{s}^{*}(t)$. Consequently, by inverse Laplace transforming Eq.(\ref{LTdiffimplicitform_bis}) we obtain 
\begin{equation}
\frac{d}{dt}\Phi _{\xi }^{(t_{a})}(t)=-\psi _{t_{a}}^{*}(t),
\label{LTdiffimplicitform2}
\end{equation}
which, by integration, yields

\begin{equation}
\Phi _{\xi }^{(t_{a})}(t)=\Phi _{\xi }^{(t_{a})}(0)-\int_{0}^{t}dt_{1}\psi
_{t_{a}}^{*}(t_{1}).  \label{newone1}
\end{equation}
The initial value of the $t_{a}$-old correlation function  can be determined by
explicitly Laplace transforming Eq. (\ref{implicitform}), calculated at $t=0$%
, with respect to $t_{a}$, thereby yielding

\begin{equation}
\hat{\Phi }_{\xi }^{(s)}(0)=\hat{\Psi } ^{*}\left( s\right) +\hat{\psi }%
^{*}\left( s\right) \hat{\Phi }_{\xi }^{(s)}(0) ,  \label{newone2}
\end{equation}
which simplifies to

\[
\left[ 1-\hat{\psi ^{*}}(s)\right] \hat{\Phi}_{\xi }^{(s)}(0)=\hat{\Psi}%
^{*}\left( s\right) . 
\]
On the other hand, we know that $\hat{\Psi}^{*}\left( s\right) =\left[ 1-\hat{\psi ^{*}}(s)\right] /s$, so the initial Laplace variable is

\[
\hat{\Phi}_{\xi }^{(s)}(0)=1/s.
\]
It is obvious that the inverse Laplace transform of this last expression
yields the initial condition for the correlation function $\hat{\Phi}_{\xi
}^{(t_{a})}(0)=1$, as it must be, based on the definition of Eq. (\ref{aging}%
) when $t_{1}=t_{2}$ and $\xi=\pm 1 $, a dichotomous variable. Using this initial
condition in (\ref{newone1}), it follows that

\begin{equation}
\Phi _{\xi }^{(t_{a})}(t)=1-\int_{0}^{t}dt_{1}\psi _{t_{a}}^{*}(t_{1})=\Psi
_{t_{a}}^{*}(t),  \label{newone3}
\end{equation}
which, as we know from Section II, coincides with the exact expression for
the  $t_{a}$-old correlation function  of Ref. \cite{luck}. 
%In conclusion, the solution of Eq. (\ref{implicitform}) coincides
%with the exact expression for the aging correlation function as obtained in \cite{luck}.

%Another way to prove (\ref{implicitform}) is to use it to rederive the
%exact formula for $\hat{\Phi_{\xi}}^{(t_{a})}(u)$.
%(In my opinion here, before the new Gerardo's derivation,
%we should, before it, put back here subsection IIC)

\subsection{Approximation through iterative expansion, and truncation}

Although we established that Eq. (\ref{implicitform}) is exact, due to its
implicit nature, it cannot be used to obtain an analytic function without
carrying out a reasonable approximation. The approximation we select rests
on replacing $\Phi _{\xi }^{(t_{l})}(t)$ in the right-hand side of Eq. (\ref
{implicitform}) with the function $A^{(t_{a})}(t)$, which denotes a
correlation function of uncertain age, which is, however, younger than the $%
t_{a}$-old correlation function defined by Eq. (\ref{implicitform}). This
correlation function reads 
\begin{equation}
A^{(t_{a})}(t)=\frac{\Phi _{\xi }(t)-\Phi _{\xi }(t+t_{a})}{1-\Phi _{\xi
}(t_{a})}.  \label{uncertain}
\end{equation}
The easiest way to derive (\ref{uncertain}) is to adopt the language of
conditional probabilities. The numerator of this expression is the
probability of not finding an event between $0$ and $t$, minus the
probability of not finding an event between $-t_{a}$ and $t$. If we call $A$
the condition of no event in the interval $(0,t)$, and $B$ the condition of
at least one event in the interval $(-t_{a},0]$, then the numerator of the
right-hand side of (\ref{uncertain}) can be identified with the joint
probability $P(A,B)$. On the other hand, the denominator of (\ref{uncertain}%
) is simply $P(B)$, the probability of the event $B$ occuring, and therefore
the function $A^{(t_{a})}(t)$ can be identified, as it should, with the
conditional probabilily $P(A|B)$, namely with the probability of finding no
event between $0$ and $t$, (\emph{i.e}. it is a correlation function), given
an event between $-t_{a}$ and $0$ (\emph{i.e.} it has an age younger than $t_{a}$).

Finally, by plugging Eq. (\ref{uncertain}) into Eq. (\ref{implicitform}) we
obtain 
\begin{equation}
\Phi _{\xi }^{(t_{a})}(t)=\Psi ^{*}(t+t_{a})+(1-\Psi ^{*}(t_{a}))\frac{\Phi
_{\xi }(t)-\Phi _{\xi }(t+t_{a})}{1-\Phi _{\xi }(t_{a})}.
\label{centralresult}
\end{equation}
This expression is not exact, but it is analytic. Moreover, the form of Eq. (%
\ref{implicitform}) suggests an iterative approach, and we can therefore
refine our result, by replacing $\Phi _{\xi }^{t_{l}}(t)$ with $%
A^{(t_{a})}(t)$, after an arbitrary number of iterations. In Fig. 1 we check
the accuracy of the first-order approximation to the exact expression given
by (\ref{centralresult}), by comparing it to the exact prediction of Eq. ({%
\ref{newsol}). The curves in Fig. 1 represent a numerical treatment
of ({\ref{newsol}) with different values of $t_{a}$. In this
example $\mu =2.5$ and $T=1.5$ (as said, $\lambda =1$). The highest curve in
the figure represents the stationary correlation function, namely $%
t_{a}=\infty $. The correlation function  $\Phi _{\xi }^{\left(t_{a}\right) }(t)$ has a faster decay, with decreasing values of $t_{a}$.
Finally, the lowest curve represents the correlation function with zero age,
namely $\Psi _{t_{a}}^{*}(t)$. The symbols in Fig. 1 overlaying the
continuous curves represent the calculations using Eq. (\ref{centralresult})
with various ages. We see perfect agreement for $t_{a}\leq \langle t\rangle $
and for $t_{a}\gg \langle t\rangle $. The agreement remains good for
intermediate values, with discrepancies comparable to the numerical
round-off errors. }}

\begin{figure}[h]
\includegraphics[width=8 cm,  height=6cm]{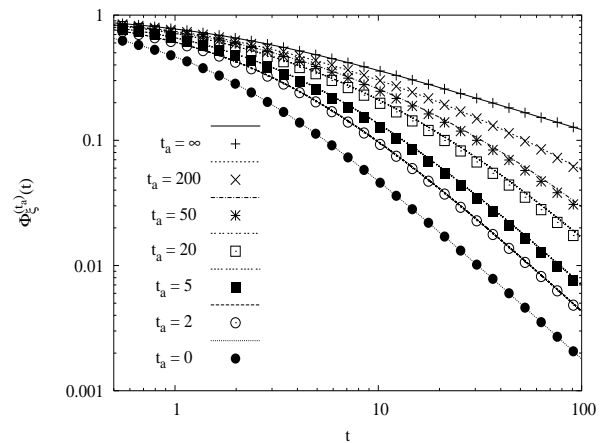}
\caption{The $t_{a}$-old correlation function, $\Phi_{\xi}^{t_a}(t)$, for different
values of $t_a$. Curves represent a numerical integration of (\ref
{newsol}) with $z=5/3$ and $T=1.5$, while dots correspond to the
approximate formula (\ref{centralresult}).}
\end{figure}

\section{concluding remarks}

The central result of this paper is the generalization of the GME discussed
in Ref. \cite{gerardo} to one of arbitrary age. Another result, closely
related to the GME of arbitrary age, is the generalization of the Onsager principle to
physical conditions of any age. The validity of this generalization of the
Onsager principle is confirmed by the fact that Eq. (\ref{exactprobability}%
), the generalized Onsager principle, yields Eq. (\ref{exact1}), and this is
shown to be equivalent to the exact prescription of Godr\`{e}che and Luck,
which is independently re-derived in Section IIC. Another interesting result
is given by Eq. (\ref{centralresult}). This is an analytical expression for $t_{a}$-old
correlation function, whose accuracy has been established using Eq. 
%(\ref{alternativeofanyage}); 
(\ref{newsol}) which is
another exact expression for the $t_{a}$-old correlation
function, determined by the Liouville-like approach of Section III.  In the
special case of non-Poisson processes where the function $\psi ^{*}(t)$ has
the form of Eq. (\ref{psistarT}), the approximate expression Eq. (\ref
{centralresult}) turns out to be very accurate. Of course there might be
non-Poisson processes where Eq. (\ref{centralresult}) is not as accurate as
in the case presented here. However, the iterative procedure discussed in
Section IVC, allows us to determine higher-order corrections, should they be
necessary for a more satisfactory treatment.

It is important to understand why Eq. (\ref{centralresult}) is not exact, in
spite of the fact that the approximation made for its derivation seems to
fit the renewal nature of the process under study, where any jump resets
memory to zero. This is a consequence of the infinite memory generated by
non-Poisson dynamics, in spite of the fact the random events reset the
memory to zero. The evaluation of the correlation function involves
probabilistic arguments, and with them the infinite memory associated with the
probabilistic treatment of non-Poisson processes.

We illuminate the meaning of the correlation function of arbitrary age by means of
the Liouville-like density approach. The observation of the process of
regression to equilibrium of the population difference corresponds to
evaluating the anti-symmetric distribution, while leaving the symmetric part
of the distribution free to evolve. If the distribution is at equilibrium,
the symmetric part corresponds to the equilibrium distribution and the
integral of the left portion of the anti-symmetric part, without back
injection, regresses to zero as the corresponding equilibrium correlation
function. For any other condition, the integral of the left portion of the
anti-symmetric part regresses to zero with an analytical expression
depending on the time at which observation begins. The regression continues
as a function of that specific initial condition while the symmetric part
keeps moving towards equilibrium independently of the population difference.
This explains why the regression to equilibrium depends on the initial
condition, of any age, with no further dependence on the bath dynamics that
keeps drifting towards equilibrium. This also explains why an emission or
absorption spectrum \cite{prl} is not stationary and changes with time. The
resonant radiation establishes a connection between the anti-symmetric and
the symmetric parts of the distribution, thereby updating observation to the
changing bath conditions.

It is worth ending this paper with some further remarks about these theoretical problems. We have built up a GME of arbitrary age, using an empirical approach. Is it possible to derive the same  GME by using a Liouville-like approach? In principle, we should use the Liouville-like picture of Section III, to derive, via contraction on the bath variables, the same GME, of arbitrary age, as that of Section II D. However, it is evident that this effort, even if we were successful, would be of limited help, for practical purposes. Suppose, for instance, that we have to study the response of the system to an external, time dependent, perturbation. Would the GME of arbitrary, but fixed, age, useful for this purpose? It is evident that it would not. In fact, the external perturbation at times different from the age of the system would produce effects departing from the more realistic approach resting on perturbing trajectories. In the specific case of the absorption spectrum of blinking quantum dots \cite{prl} the authors adopted in fact this trajectory perspective to make a theoretical prediction  that is incompatible with the perturbation of a GME of fixed age. In literature, there already exists at least the discussion of one case [16] that seems to be a natural consequence of this property. Sokolov, Blumen and Klafter \cite{externalfield} derived an exact density equation to describe a sub-diffusion process. This equation corresponds to brand new initial conditions. This means a condition where $t_{a} = 0$. Then, these authors perturbed this equation with a time dependent field, and they found that the theoretical result conflicts with the behavior  of the CTRW under the influence of the same perturbation.
      This is so because the time dependent perturbation corresponds to additional observation, taking place  at different  time values, none of them coinciding with the observation time, but the perturbation at $t= 0$. This is true, whatever the observation time is, either $t_{a} = 0$, as in Ref. \cite{externalfield}, or $\infty > t_{a} >0$, a condition requiring the GME of this paper. Regardless of the observation time that we assign to the GME, it is impossible to make the GME prediction identical to the CTRW prediction, if we require the perturbation to remain external to the system. The concept of perturbation itself turns out to be inadequate to study non-Poisson processes, regardless of its intensity. Thus, the results of this paper, in addition to shedding light into those of Ref. \cite{externalfield}, imply a violation of the linear response theory. The only possible way to make the density compatible with the trajectory picture is to make the external perturbation become a part of the system to study. This means that we have to build up a totally new, field-dependent, GME, along the lines of Ref.  \cite{prl}. This sets a limit on the applicability of the GME of arbitrary age found in this paper. However, this result seems to support the conclusion that the trajectory-density conflict, revealed by Bologna, Grigolini and West \cite{chemphys} might be a consequence of the aging properties emerging from non-Poisson renewal process.

\end{document}